%% file: PMB_template_modified.tex
\documentclass[12pt]{iopart}

\usepackage{graphicx} 
\usepackage{tabularx}
\usepackage{iopams}
\usepackage{url}
\usepackage[colorinlistoftodos]{todonotes}
\graphicspath{ {./images/} }
\usepackage{float} 
\usepackage{xcolor}
\usepackage{natbib}
\usepackage{subcaption}
\usepackage[ruled,vlined]{algorithm2e}
\usepackage{tikz}
\usetikzlibrary{arrows.meta,positioning,shapes.geometric}

\bibpunct[, ]{(}{)}{,}{a}{}{,}%
\begin{document}

\title[Extending gPET for multi-layer DOI detectors]%
{Extending gPET for Multi-Layer PET Simulation}

\author{Satzhan Sitmukhambetov$^1$, Junwei Du$^2$, Mingwu Jin$^1$, Yujie Chi$^1$}

\address{$^1$Department of Physics, The University of Texas at Arlington, Arlington, Texas, United States\\
$^2$Department of Biomedical Engineering, University of California at Davis, Davis, California, United States}
\ead{Yujie.Chi@uta.edu}
\vspace{10pt}

\begin{abstract}
\textbf{Objective.} Depth-of-interaction (DOI) encoding is an effective strategy for reducing parallax error and preserving spatial resolution in positron emission tomography (PET), particularly in compact small-animal scanners. To enable efficient simulation-driven design of DOI-capable systems, we extend the GPU-accelerated Monte Carlo toolkit \textit{gPET} to support flexible multi-layer detector geometries. \textbf{Approach.} The original three-level hierarchical detector model in \textit{gPET} (panel-module-crystal) was expanded by introducing an intermediate ``layer'' level, enabling parameterized modeling of stacked scintillator architectures. The photon-transport algorithm was correspondingly updated to sample interactions across multiple layers and panels while preserving GPU-efficient memory usage. The framework was validated using three scanner configurations: a conventional single-layer ring (H2RSPET-1CL), an aligned split-layer design (H2RSPET-1CL-split), and an offset dual-layer design (H2RSPET-2CL). System performance was evaluated following NEMA NU4-2008 protocols using sensitivity, spatial resolution, and Derenzo phantom simulations with CASToR-based Maximum Likelihood-Expectation Maximization (MLEM) reconstruction. \textbf{Main Results.} The H2RSPET-1CL and H2RSPET-1CL-split configurations produced statistically identical hit distributions, while H2RSPET-2CL design exhibited offset interaction patterns. Sensitivity of the H2RSPET-2CL design remained comparable to the H2RSPET-1CL system, generally within $\sim$2-5\% difference, while radial spatial resolution improved substantially (0.8-1.6~mm vs.\ 1.0-4.2~mm from the center to a 50~mm radial offset). Derenzo phantom simulations also demonstrated improved rod separation and contrast recovery for the H2RSPET-2CL design. Runtime performance remained essentially unchanged between configurations. \textbf{Significance.} The extended \textit{gPET} framework enables fast, flexible simulation of multi-layer PET detectors and supports efficient optimization of DOI-enabled PET system designs.
\end{abstract}

\input{subsections/intro}
\input{subsections/methods}
\input{subsections/results}

\input{subsections/discussion}

\input{subsections/conclusion}

\input{subsections/acknowledgment}

\bibliography{references}
\end{document}

%% file: subsections/intro.tex
\section{Introduction}

Positron emission tomography (PET) is a quantitative molecular imaging modality widely used in both preclinical and clinical applications to visualize radiotracer distributions noninvasively. Continued advances in detector technology and system geometry are enabling PET scanners with higher sensitivity and improved spatial resolution, but they also increase the complexity of scanner design and performance optimization. Because building and iterating on physical prototypes is expensive and time-consuming, Monte Carlo (MC) simulation has become a standard tool for predicting scanner performance prior to optimizing scanner design and enabling fair comparisons across alternative configurations.

A number of MC simulation toolkits have been developed for PET. One well known example is the GATE toolkit \citep{Jan2004GATE}, which provides flexible, high-fidelity modeling and has been widely adopted by the community. However, the earlier GATE is CPU-based, and comprehensive evaluations, especially those requiring repeated simulation across many candidate scanner designs, remain computationally intensive even when using multithreading and distributed computing. This limitation becomes increasingly consequential for modern systems with long axial fields-of-view and detectors with fine crystal pitch.

With the growing availability of GPU computing and the increasing computational demands of PET system development, several GPU-accelerated PET simulation tools have been introduced \citep{Chi2025_Review}, including GGEMS-PET \citep{Ma2019GGEMSPET}, gPET \citep{Lai2019gPET}, MCGPU-PET \citep{Herraiz2024MCGPU}, and UMC-PET \citep{Galve2024UMC}, among others. These tools substantially reduce simulation time and thus enable more practical iterative optimization. Nevertheless, GPU-based simulation support for multi-layer scintillator detectors, a key detector architecture for encoding depth-of-interaction (DOI) information, remains limited. In particular, some existing tools either do not explicitly support parameterized detector modeling or rely heavily on phase-space descriptions, which can restrict their utility when detailed multi-layer detector studies are required. While gPET introduced a parameterized and GPU-memory-efficient geometry representation, it currently supports only single-layer detector configurations. More recently, UMC-PET explicitly added multi-layer detector modeling on GPUs, but its voxelized geometry representation can become memory demanding for detectors with fine structural features such as thin crystal layers, reflectors, or inter-layer gaps.

These limitations are increasingly important because, as PET systems push toward higher spatial resolution, the impact of DOI uncertainty becomes more significant and can limit achievable reconstructed resolution, especially in small-animal PET, where the smaller detector ring diameter amplifies parallax error. A range of detector architectures has been developed to encode DOI information and reduce parallax effects, including dual-ended readout detectors \citep{Du2021DualEnded}, multi-layer scintillator designs \citep{Bouziri2021Phoswich}, and monolithic scintillators with machine-learning-based positioning \citep{Sanaat2020_MonolithicML}. Among these approaches, multi-layer crystal detectors with single-ended readout are attractive for compact, cost-sensitive systems because they offer DOI encoding with comparatively simple hardware and straightforward integration, making them well suited for desktop-scale small-animal PET scanners. Consequently, a fast and flexible GPU-based simulation tool that can efficiently model multi-layer crystal geometries is valuable for accelerating scanner development and enabling systematic DOI-related trade studies.

In this work, we extend the gPET framework to flexibly support multi-layer detector configurations while retaining its memory-efficient parameterized geometry representation. The proposed extension allows users to define multiple radially stacked crystal layers with independent offsets and provides layer-resolved list-mode output suitable for DOI analysis. To demonstrate and validate the new capabilities, we simulated three representative small-animal PET configurations: (i) a conventional single-layer ring, (ii) a split two-layer design, and (iii) a two-layer offset design inspired by prior dual-layer DOI studies \citep{Yang2008_PrototypeDOI, Thompson2013DualLayer,Zhang2022DualLayer}. System performance was evaluated following NEMA NU 4 metrics. The results confirm that the proposed gPET extension accurately models multi-layer detector geometries and reproduces expected DOI-related improvements in spatial resolution, enabling efficient simulation-driven design exploration for DOI-capable PET systems.

%% file: subsections/methods.tex
\section{Methods}

\subsection{Extension of gPET for multi-layer PET detector simulation}
\label{subsec:upgrade_gpet}

In the initial development of gPET \citep{Lai2019gPET}, a PET detector was described using a hierarchical parametrization consisting of panels, modules, and crystal units. User inputs include panel position and orientation in the global coordinate, panel size, module size and spacing, crystal size and spacing, as well as material types and densities for the crystal and inter-crystal interval. Based on a single panel definition, the user can construct the complete PET scanner either in the repetition mode by specifying the number of repetitions and the rotational angle for panel placement, or in the non-repetition mode by explicitly listing the configuration of each panel sequentially. With this structure, gPET assigns each interaction event a unique crystal, module, and panel index, and efficiently transports a gamma escaping from the phantom to its corresponding panel via line–plane intersection calculations. The subsequent interaction within the detector is simulated by transforming the gamma’s global coordinates into the panel’s local frame and performing step-by-step MC transport in the crystal array. This hierarchical design also allows the user to flexibly choose among different single event  readout levels including crystal-level, module-level, or panel-level, depending on the intended detector modeling or data-processing application.

\begin{figure}[hbt!]
  \centering
  \includegraphics[width=\linewidth]{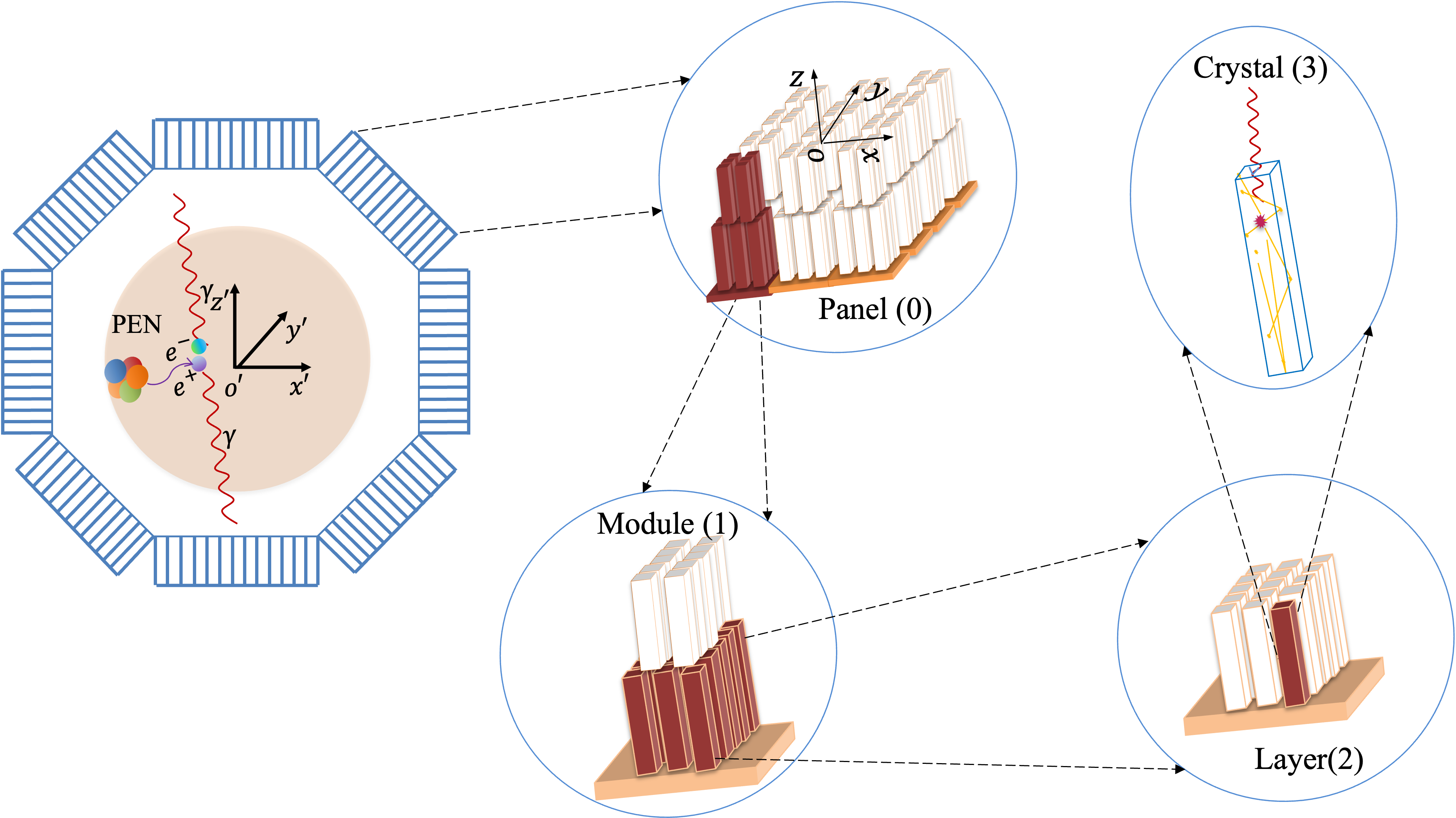}
  \caption{Illustration of the extended hierarchical description of a PET detector.
Compared with the original three-level geometry used in gPET (panel, module, crystal)\citep{Lai2019gPET}, the extended framework introduces a new layer level between the module and crystal for a four-level hierarchy: panel (level 0, highest), module (level 1), layer (level 2), and individual crystal (level 3, lowest).}
  \label{fig:illustration}
\end{figure}

To extend gPET to incorporate the multi-layer configuration, we preserved the original hierarchical geometry description in order to maintain the GPU-friendly memory usage and high computational throughput of the code. As multi-layer PET detector designs are generally organized at the module level, we introduced a new level, termed the “layer” index, inserted between the previous module and crystal levels (Fig. \ref{fig:illustration}). Correspondingly, the detector configuration file was expanded to include a new parameter, “the number of layers”, allowing the user to specify the total number of crystal layers within a module. The material type and density entries were modified so that the crystal and inter-crystal interval materials can be provided sequentially for each layer. Similarly, below the module size and spacing definitions, the user can now specify the crystal size, spacing, and offset for each layer in order. All remaining configuration fields were preserved without modification, including the repetition and non-repetition modes used to construct the full scanner geometry. Corresponding updates were also made to the geometry reader so that gPET now labels each interaction event with crystal, layer, module, and panel indices. Meanwhile, we kept the single event output at the crystal, module, and panel levels.

In addition to the geometry extension, we also updated the gamma transport procedure inside the panel’s local coordinate system. In the previous implementation, once a gamma reached a panel, it was transported step by step until it either exited the panel or fell below the cutoff energy. During transport, the Woodcock tracking technique was used to simplify free-path sampling in the presence of heterogeneous material compositions, and the repetitive crystal structure facilitated rapid determination of material types and crystal indices for interaction sampling and event recording. In the multi-layer extension, the same free-path sampling strategy is retained. The key addition is a layer-selection step: based on the gamma’s depth coordinate in the panel’s local frame, the appropriate crystal layer is identified and the corresponding crystal parameters are loaded for material identification and interaction sampling.

\subsection{Configurations to validate gPET extension}
\label{subsec:validation}

\begin{figure}[hbt!]
  \centering
  \includegraphics[width=\linewidth]{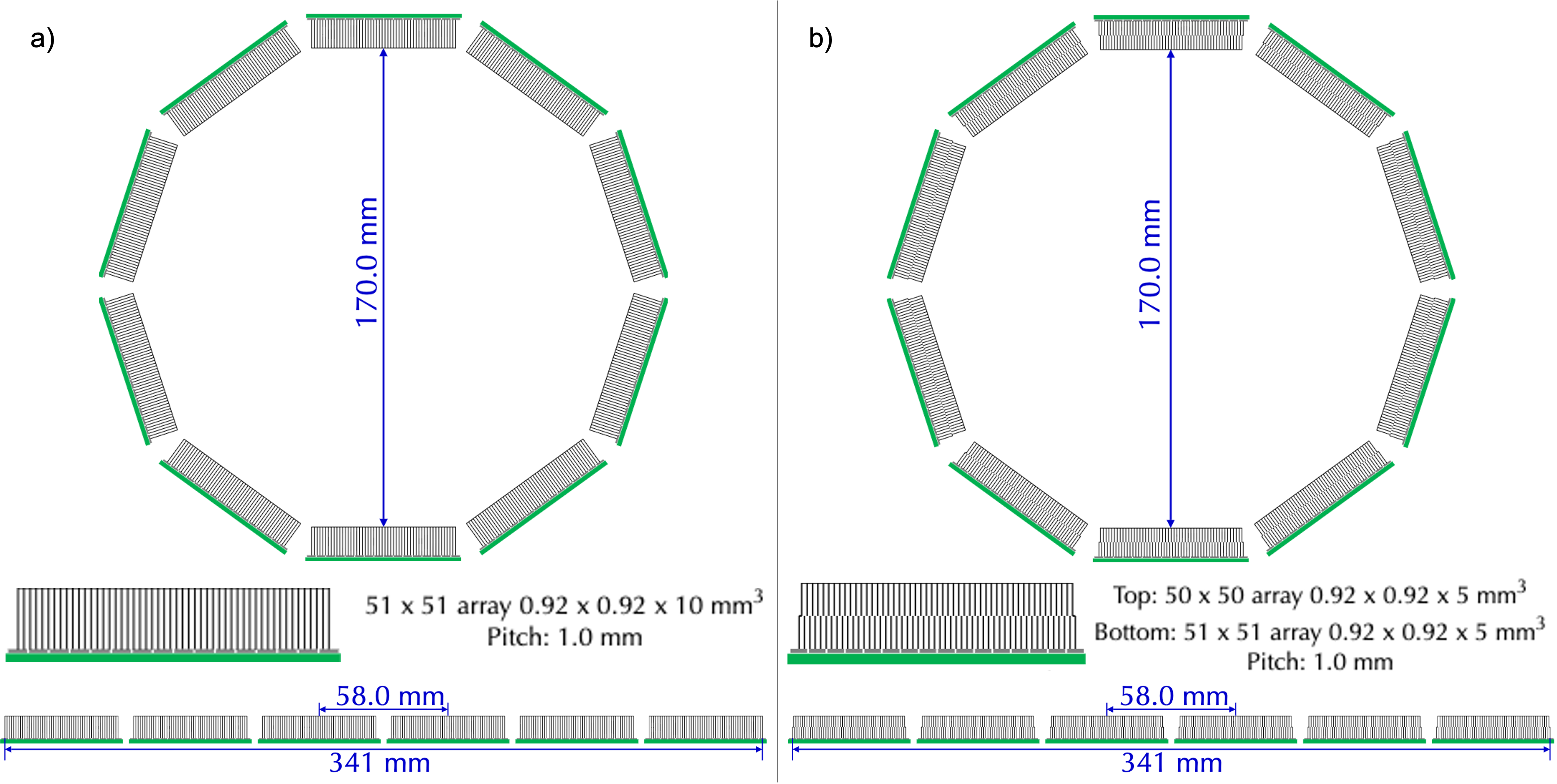}
  \caption{Geometry illustration of (a) H2RSPET-1CL PET and (b) H2RSPET-2CL PET. The two scanners have the same diameter, axial length, and pitch size, except H2RSPET-2CL PET is a two-layer design with a half-pitch offset between the two layers.}
  \label{fig:h2rspet}
\end{figure}

To validate the extended gPET framework, three PET scanner configurations were simulated, as illustrated in Fig.~\ref{fig:h2rspet}. The first configuration is a single-layer ring, referred to as H2RSPET-1CL (Fig.~\ref{fig:h2rspet}(a)). It consists of ten identical detector panels arranged concentrically at a radius of 85~mm. Each panel has a physical dimension of $1.0 \times 5.1 \times 34.1$~cm$^3$ (radial $\times$ tangential $\times$ axial). Within each panel, six detector modules are implemented axially, each containing a $51 \times 51$ array of 10~mm-deep LYSO crystals with an approximately 1.0~mm pitch (0.92~mm crystal + 0.08~mm air gap), separated by a 7~mm inter-module gap. The second configuration is an aligned two-layer ring, referred to as H2RSPET-1CL-split, in which each original 10~mm crystal from H2RSPET-1CL is replaced by two radially connected 5~mm crystals. The third configuration is an offset two-layer ring, referred to as H2RSPET-2CL (Fig.~\ref{fig:h2rspet}(b)). H2RSPET-2CL shares the same panel geometry, number of panels, and number of modules per panel as H2RSPET-1CL. The difference lies inside each module, where two layers of 5~mm-long crystals are stacked radially: the first (inner) layer contains a $50 \times 50$ crystal array and the second (outer) layer contains a $51 \times 51$ array, with a half-crystal offset in the axial direction between layers.

In gPET simulations, the scintillator material for all three scanner configurations was set to LSO/LYSO with a density of $7.4$~g/cm$^3$. The origin of the global coordinate system was placed at the radial and axial center of the scanner ring, with the $x$-axis pointing rightward, the $y$-axis upward, and the $z$-axis outward, following the scanner view shown in Fig.~\ref{fig:h2rspet}. Hit distribution was recorded for each scanner configuration and was compared in the crystals profiled locally and globally.


\subsection{Performance evaluation for single-layer and two-layer configurations}
\label{subsec:application}

With the validation of the multi-layer extension of gPET, we next moved to a systematic comparison of imaging performance between H2RSPET-2CL and H2RSPET-1CL via sensitivity profiles, spatial resolution analysis, and small rod visibility analysis.

We followed the NEMA NU4-2008 standard for both sensitivity and spatial resolution analysis using an $^{18}$F point source. The source contained $1.6 \times 10^{8}$ atoms, corresponding to an initial activity of about 
$1.70 \times 10^{-2}\,\mathrm{MBq}$.
A decay simulation time of $120\,\mathrm{s}$, a coincidence timing window of $3\,\mathrm{ns}$, and an energy resolution of $16\%$ were applied in all simulations. Positron range simulation was disabled, and annihilation was assumed to occur at the positron emission site. 

Sensitivity $S(r,E)$ as a function of location $r$ and energy $E$ was computed over radial offsets ranging from $0$ to $70\,\mathrm{mm}$ in $5\,\mathrm{mm}$ increments and axial offsets ranging from $0$ to $170\,\mathrm{mm}$ in $10\,\mathrm{mm}$ increments. Energy windows of $150$-$700\,\mathrm{keV}$, $250$-$700\,\mathrm{keV}$, $350$-$700\,\mathrm{keV}$, and $450$-$700\,\mathrm{keV}$ were evaluated. Sensitivity difference between the two scanner configurations was then evaluated as $\Delta_{\%}^{\mathrm{sym}}(r,E)=200\,\frac{S_{\mathrm{1CL}}(r,E)-S_{\mathrm{2CL}}(r,E)}{S_{\mathrm{1CL}}(r,E)+S_{\mathrm{2CL}}(r,E)}$.

In spatial resolution simulation, an air phantom with a voxel size of $0.1 \times 0.1 \times 0.1$~mm$^3$ and a dimension of $2.1\times2.1\times2.1\ \mathrm{cm^3}$ was used. In each simulation, the same \textsuperscript{18}F point source was placed at the center voxel, and a background activity was set for the rest of the voxels. The background noise was approximately 10\% of the peak intensity of the reconstructed point source voxel. An energy window of 250-700~keV was used for coincidence selection. We used the open source CASToR (Customizable and Advanced Software for Tomographic Reconstruction) package (version 3.1.1) \citep{merlin2018castor} to reconstruct the 3D images from gPET recording. The list-mode coincidence events from gPET were converted to CASToR version of list-mode events. Twelve iterations of the 3D Maximum Likelihood Expectation Maximization (MLEM) algorithm (non-Time of Flight) were used for all reconstructions. The Full Width at Half Maximum (FWHM) of the output profile along the radial, tangential, and axial directions was then calculated following the NEMA NU4-2008 standard.

A Derenzo phantom was used for the visibility analysis. The phantom consists of six triangular sectors arranged to form a hexagonal geometry. Each sector contains an array of 2~cm length rods with the same diameter, and the center-to-center spacing between adjacent rods equals twice the rod diameter. The rod diameters for the six sectors are 0.3, 0.4, 0.5, 0.625, 0.75, and 1.0~mm, and the corresponding rod counts for each section are 6, 10, 15, 21, 28 and 36, respectively. Each rod was assigned a uniform radioactive \textsuperscript{18}F activity density of $9.21\times10^{-5}\ \mathrm{MBq/mm^3}$ and a total decay simulation time of $1200\ \mathrm{s}$ was used to generate sufficient decay events. Except for the rods, the rest of the phantom was set to be a water box covering a volume of $4.025 \times 4.025 \times 4.025$~cm$^3$. The reconstructions were conducted with 12 iterations of the MLEM algorithm with $161 \times 161 \times 161$ grid and $0.25 \times 0.25 \times 0.25$~mm$^3$ voxels.

\subsection{Computational Performance}
To confirm that the multi-layer extension does not introduce substantial computational burden, we compared wall-clock runtimes for the single-layer, split-layer, and dual-layer scanner configurations using the Derenzo activity distribution described in Section~\ref{subsec:application}. For each configuration, the full MC simulation was run 10 times on a single NVIDIA TITAN Xp GPU (12~GB; NVIDIA driver 510.47.03; CUDA 11.6) to obtain the average runtime.

%% file: subsections/results.tex
\section{Results}

\subsection{Validation of gPET multi-layer extension}
\label{res:internal_validation}
\begin{figure}[hbt!]
  \centering
  \includegraphics[width=1.0\linewidth]{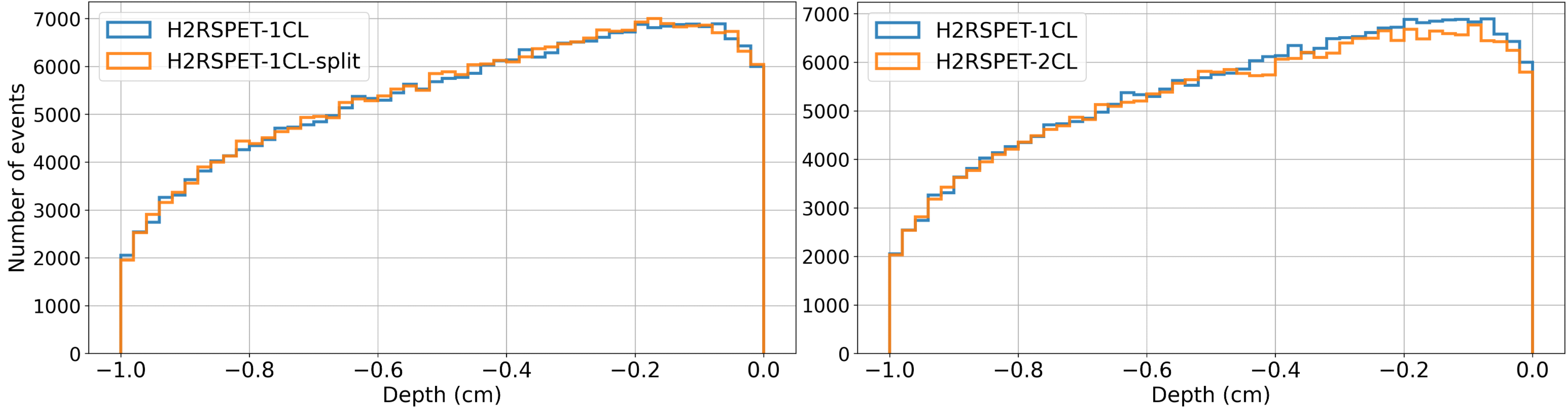}
  \caption{Histogram of hit depth events in the H2RSPET detectors: (left) H2RSPET-1CL versus H2RSPET-1CL-split, and (right) H2RSPET-1CL versus H2RSPET-2CL. The $x$-axis (cm) denotes the depth of interaction (DOI) in the panel's local coordinate frame.}
  \label{fig:doihists}
\end{figure}

Fig.~\ref{fig:doihists} shows the comparisons of the DOI distributions of hit events between the H2RSPET-1CL and H2RSPET-1CL-split configurations, and between the H2RSPET-1CL and H2RSPET-2CL configurations. As expected, the hit distributions along the depth in the panel’s local coordinate system agree closely between the H2RSPET-1CL and H2RSPET-1CL-split configurations, as they share exactly the same detector geometries except for the layer split. In contrast, the H2RSPET-2CL scanner exhibits fewer hit events in the shallow DOI region (0–4.5~mm), while the deeper DOI region matches well with H2RSPET-1CL. This reduction is consistent with the fact that the top layer of H2RSPET-2CL contains fewer crystals per module ($50 \times 50$ vs.\ $51 \times 51$), resulting in a lower total crystal volume for potential interactions. The observed DOI-dependent behavior thus confirms that the extended gPET framework produces physically consistent and reasonable gamma-transport results for multi-layer detector geometries.

\begin{figure}[hbt!]
  \centering
  \includegraphics[width=1.00\linewidth]{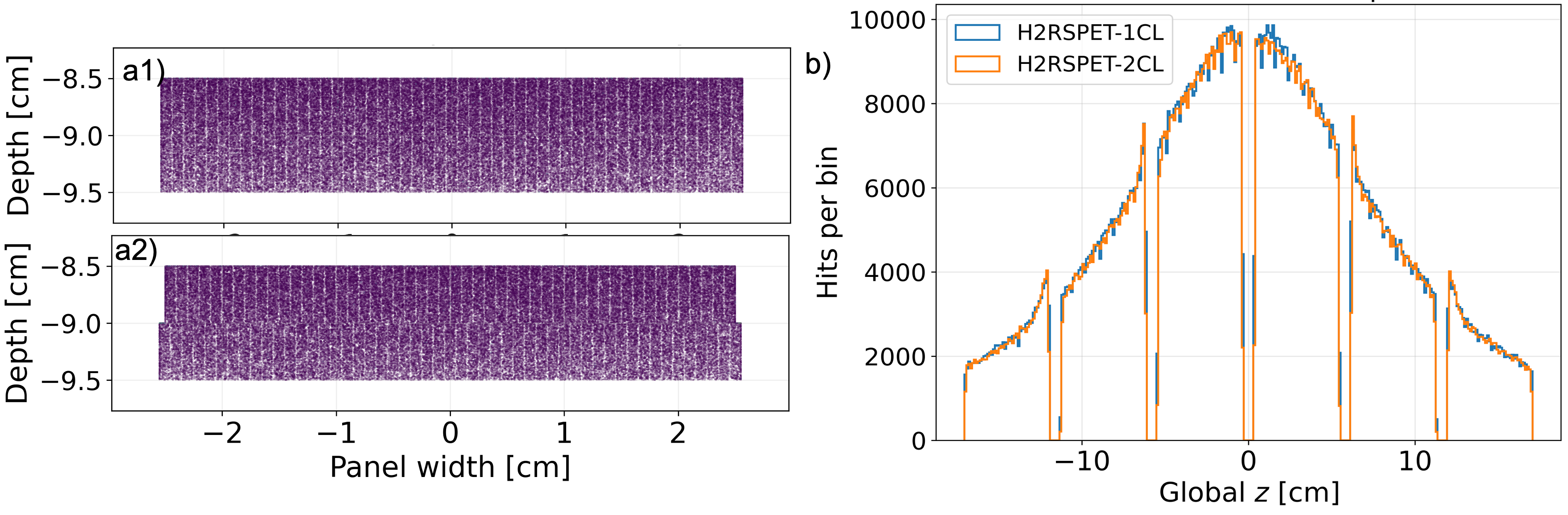}
  \caption{Scatter plot within panel's $x$ and $y$ view for a1) the H2RSPET-1CL split-layer configuration and a2) the H2RSPET-2CL scanner. b) Axial distributions of hit events along the global $z$ direction for H2RSPET-1CL and H2RSPET-2CL with a central point source. }
  \label{fig:hits}
\end{figure}

Fig.~\ref{fig:hits}(a1) and (a2) show the scatter distributions of hit events in the global $x$-$y$ plane for the H2RSPET-1CL-split and H2RSPET-2CL scanners, respectively, for a representative module whose crystal-depth direction is aligned with the global $-y$ axis. The hits are integrated along the global $z$ (axial) direction. As expected, H2RSPET-1CL-split exhibits no discernible difference from the nominal single-layer geometry, whereas H2RSPET-2CL displays a clear lateral shift associated with the offset of the second crystal layer. Fig.~\ref{fig:hits}(b) presents the axial distributions of hit events along the global $z$ direction. With the point source positioned at the scanner center, both H2RSPET-1CL and H2RSPET-2CL produce symmetric axial profiles. Collectively, these results provide strong evidence that the extended gPET framework accurately models multi-layer detector geometries, thereby establishing a reliable foundation for subsequent performance comparisons between detector designs.

\subsection{Performance evaluation for single-layer and two-layer configurations}
\begin{figure}[hbt!]
  \centering
  \includegraphics[width=0.85\linewidth]{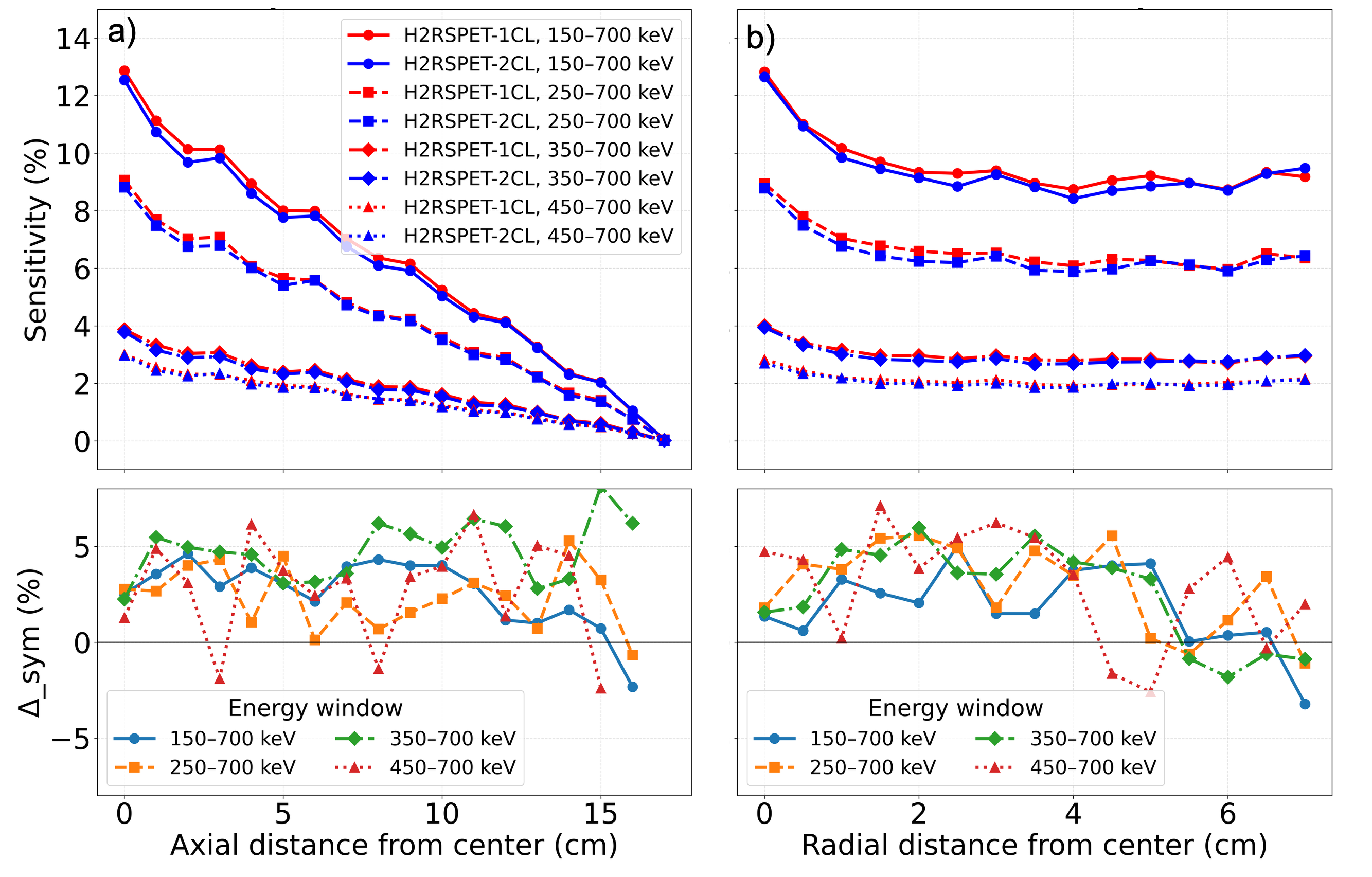}
  \caption{Sensitivity profiles and percent differences for H2RSPET-1CL and H2RSPET-2CL scanner configurations along (a) axial and (b) radial directions.}
  \label{fig:sensitivity_profiles}
\end{figure}

Fig.~\ref{fig:sensitivity_profiles} compares the axial and radial sensitivity profiles for the H2RSPET-1CL and H2RSPET-2CL configurations. Sensitivity decreases as the distance from the center increases along both directions. The 1CL configuration exhibits a slightly higher absolute sensitivity than the 2CL configuration across most positions and energy windows. Percent-difference analysis reveals this difference is generally within 2-5\%; however, larger fluctuations occur at extreme axial positions due to reduced counting statistics. The discrepancy can be partially attributed to the 1.94\% reduction in detector volume in the 2CL design compared to the 1CL ($50 \times 50$ vs $51 \times 51$ crystals per module).  

\begin{figure}[hbt!]
  \centering
  \includegraphics[width=0.8\linewidth]{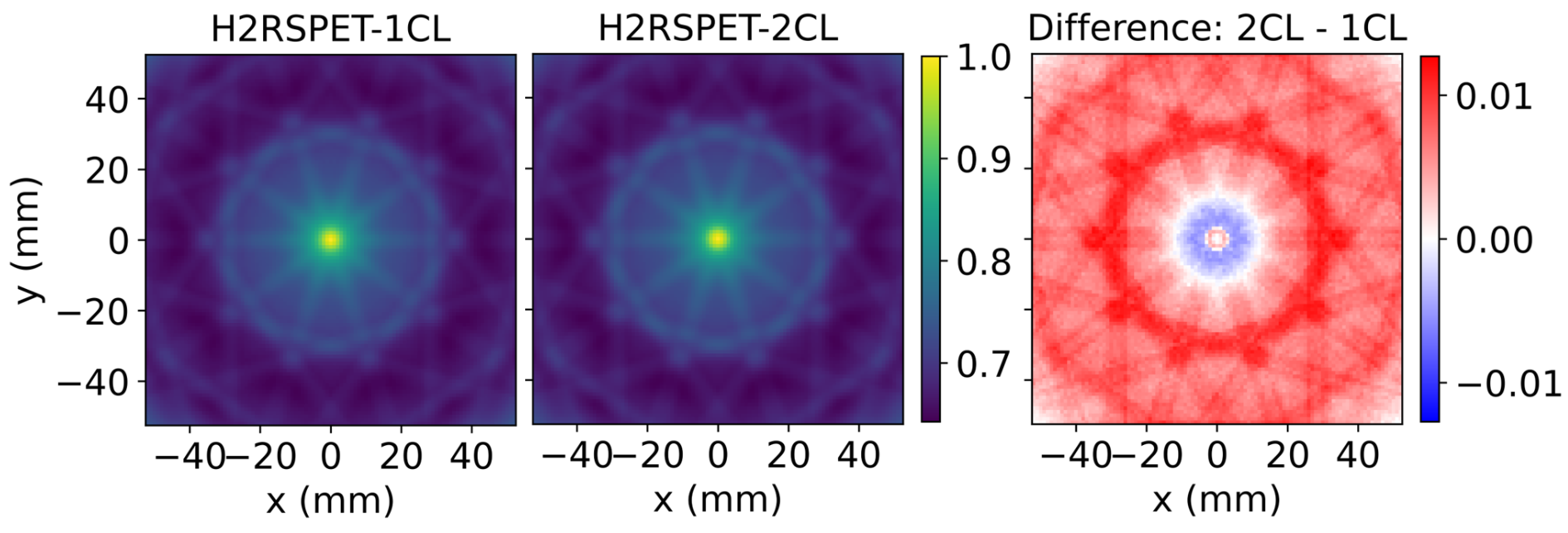}
  \caption{Central axial slices of the normalized sensitivity distributions for H2RSPET-1CL (left) and H2RSPET-2CL (middle). The right panel shows the difference map (2CL$-$1CL), highlighting spatial regions where the two configurations differ in relative sensitivity.}
  \label{fig:sensitivity_map}
\end{figure}

A representative sensitivity map for the central phantom slice obtained from the H2RSPET-1CL and H2RSPET-2CL scanners is shown in Fig.~\ref{fig:sensitivity_map}. Both configurations exhibit high sensitivity near the center with a radial fall-off, and the angular modulation reflects the ten-panel geometry of the system. With both maps normalized to their respective maxima, the difference image shows that the offset crystal layer in H2RSPET-2CL yields a small redistribution and flattening of the profile, likely due to the staggered crystal arrangement reducing inter-crystal gaps. Except for this localized improvement, the overall sensitivity patterns remain similar between the two designs.

\begin{figure}[hbt!]
  \centering
  \includegraphics[width=\linewidth]{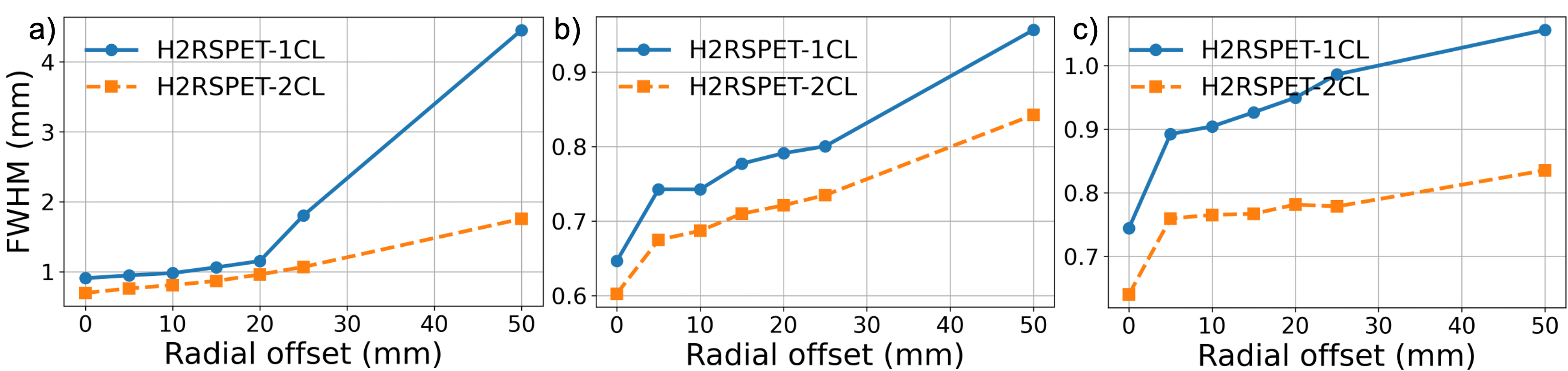}
  \caption{Spatial resolution for the reconstructed point-source images as a function of radial offset for both H2RSPET-1CL and H2RSPET-2CL scanners. At each source position, the FWHM profile is measured along (a) radial, (b) axial, and (c) tangential direction in millimeters.}
  \label{fig:fwhm_all}
\end{figure}

Fig. \ref{fig:fwhm_all} shows the spatial resolution performance of H2RSPET-1CL and H2RSPET-2CL as a function of radial offset. The axial (Fig.~\ref{fig:fwhm_all}(b)) and tangential (Fig.~\ref{fig:fwhm_all}(c)) FWHM curves exhibit modest but consistent improvements for H2RSPET-2CL configuration. The most pronounced difference appears in the radial FWHM (Fig.~\ref{fig:fwhm_all}(a)). Along this direction, H2RSPET-2CL curve increases moderately, from 0.8 mm at the center to 1.6 mm at a 50 mm radial offset, compared to 1.0 and 4.2 mm for H2RSPET-1CL at the corresponding locations. These results show that the two-layer offset design effectively suppresses parallax-induced blurring and preserves spatial resolution, particularly in regions farther from the scanner center. This study also added the evidence that DOI segmentation primarily mitigates parallax effects along the radial direction.

\begin{figure}[hbt!]
  \centering
  \includegraphics[width=0.8\linewidth]{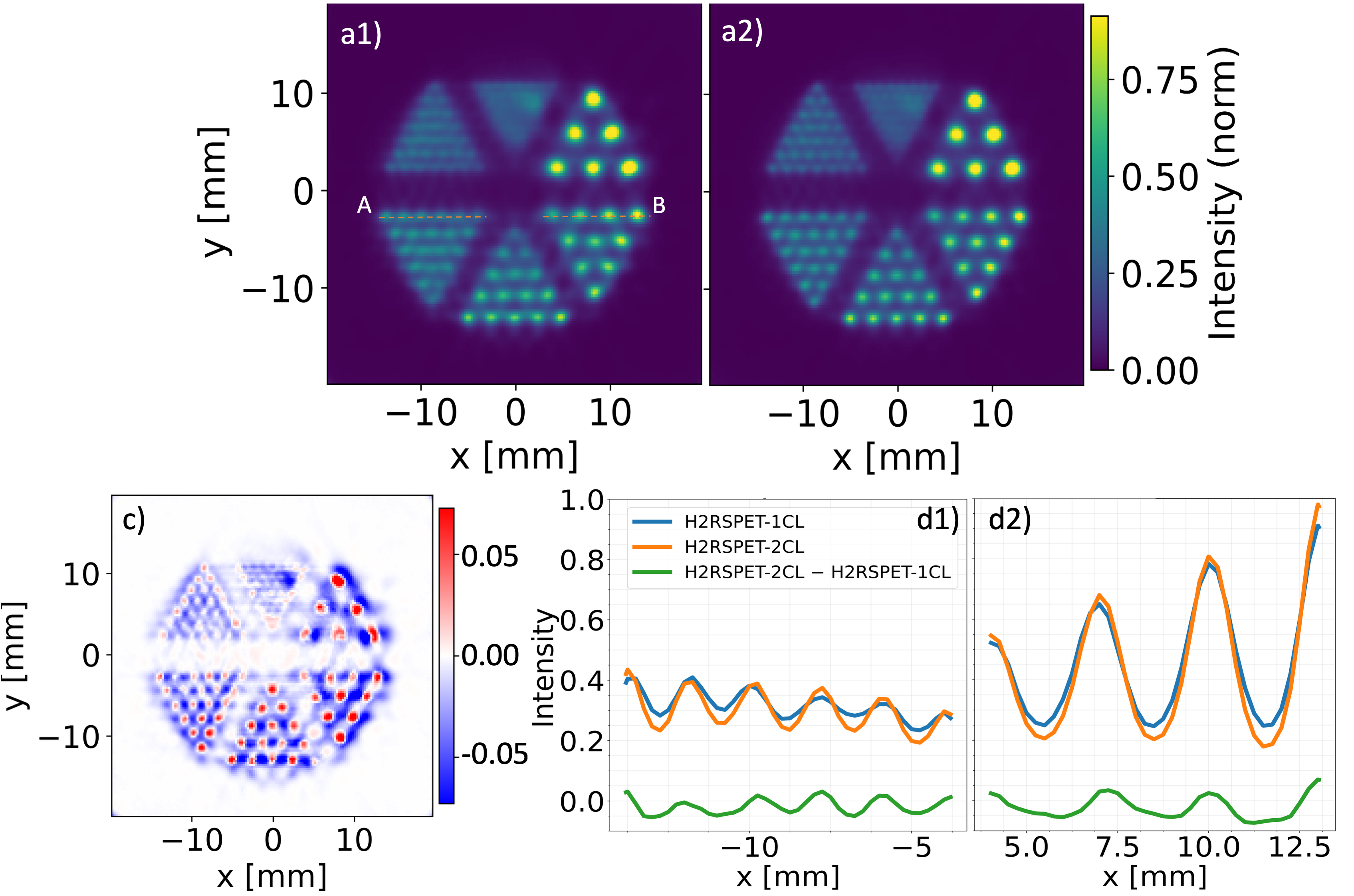}
  \caption{Reconstructed Derenzo phantom results comparing H2RSPET-1CL and H2RSPET-2CL configurations. (a1) and (a2) Normalized intensity maps for the 1CL and 2CL scanners, respectively. (c) Intensity difference map (2CL $-$ 1CL), where positive values indicate regions with higher signal in the two-layer design and negative values indicate reduced signal. (d1) and (d2) Line profile comparisons along the A and B lines indicated in (a1), showing local differences between the two configurations.}
  \label{fig:derenzo116}
\end{figure}

Fig.~\ref{fig:derenzo116} shows reconstructed Derenzo phantom images for H2RSPET-1CL and H2RSPET-2CL, together with their difference map and line-profile analysis. As shown in Figs.~\ref{fig:derenzo116}(a1) and (a2), both configurations resolve the larger rods, while the intermediate rods appear slightly blurred in the single-layer reconstruction. The smallest rods remain challenging for both systems; however, the two-layer design still exhibits observable improved contrast and rod separation in these regions. In the intensity difference map illustration (Fig.~\ref{fig:derenzo116}(3)), red regions correspond to locally higher recovered rod intensity in H2RSPET-2CL, whereas blue regions indicate slightly lower background levels. From it, improved contrast recovery for H2RSPET-2CL is observed. Line profiles along A and B (Figs.~\ref{fig:derenzo116}(d1) and (d2)) show similar spatial resolution overall, with H2RSPET-2CL demonstrating higher peak-to-valley contrast in the selected rod groups, consistent with the contrast improvements visible in the images.

\subsection{Computational Performance}

As summarized in Table~\ref{tab:runtimes}, the runtimes for the Derenzo phantom simulations using the 3 configurations mentioned above are statistically similar. This demonstrates that the multi-layer extension of gPET does not introduce any measurable increase in simulation time.

\begin{table}[hbt!]
  \centering
  \caption{Clock runtimes for the single-layer, split-layer, and two-layer scanner configurations using the Derenzo phantom rods decay time of 12s}
  \begin{tabular}{ccccc}
    \hline
    Configuration & Runs & Mean time$\pm$ std. [s] & Hit events & Coincidence events \\
    \hline
    H2RSPET-1CL & 10 & 41.47$\pm$0.12 & 2671908$\pm$2261 & 378152$\pm$1289 \\
    H2RSPET-1CL-split & 10 & 41.55$\pm$0.06 & 2673328$\pm$2264 & 378866$\pm$544 \\
    H2RSPET-2CL & 10 & 41.25$\pm$0.11 & 2620170$\pm$3093 & 363893$\pm$695 \\
    \hline
  \end{tabular}
  \label{tab:runtimes}
\end{table}

%% file: subsections/discussion.tex
\section{Discussion}

DOI has long been recognized as a critical factor in maintaining high spatial resolution and quantitative accuracy, particularly for regions far away from the scanner center in both clinical and preclinical PET systems. In small-animal PET, sub-millimeter resolution is required to visualize the subregions of the rodent brain and the small lesions and to mitigate partial-volume effects \citep{shukla2006positron,zeng2024depth,soret2007partial}. Even at the larger, few-millimeter resolutions typical for clinical brain or organ-dedicated PET, DOI-related parallax errors continue to degrade off-axis resolution. Consequently, detector designs capable of resolving DOI are essential to reduce parallax blurring and partial-volume bias, thus improving the accuracy of metabolic measurements relevant to therapy planning and response assessment. In this context, the extension of gPET framework provides a fast and flexible simulation environment for studying DOI-aware detector architectures. Its hierarchical, parametrized geometry enables GPU-efficient, high-throughput modeling of multi-layer configurations with user-selectable output granularity. In addition, the event-level outputs, including explicit crystal, layer, module, and panel identifiers, support development of reconstruction methods that relax the assumption of a perfectly cylindrical detector ring \citep{Zhang2013_LORRecon} and accommodate generalized multi-layer architectures such as three- and four-layer stacks \citep{Tsuda2004_FourLayerDOI}.

While several hardware strategies exist to mitigate parallax error, including dual-ended readout, phoswich detectors, and monolithic crystals \citep{Liu2022}, the two-layer offset configuration examined in this study represents a practical discrete DOI approach that provides two depth bins through a comparatively simple hardware modification, avoiding the added complexity associated with dual-ended or phoswich implementations \citep{Du2021_DualEnded,Bouziri2021Phoswich}. gPET simulations demonstrate that such a dual-layer design can substantially improve off-center spatial resolution, with gains up to 2.6 times for the configurations evaluated in this study. Although our examples focus on small-animal geometries, the underlying parallax mechanisms and the associated DOI benefits can be readily translated to organ-dedicated and brain PET systems.

The simulated spatial-resolution behavior observed in this study is consistent with experimental reports that demonstrated improved off-axis resolution when DOI information is available. \cite{Thompson2013DualLayer} reported that, in a small-animal PET system, the radial resolution decreased to $3.4$ mm FWHM at a 16 mm radial offset for a single-layer detector, whereas a two-layer design sustained a markedly better resolution of $1.83$ mm FWHM. Similarly, \cite{Yang2008_PrototypeDOI} showed that increasing DOI sampling improved the reconstructed resolution from $1.6$ mm to $0.9$ mm FWHM. Although our simulations were not intended to replicate any specific prototype system, the dual-layer H2RSPET-2CL configuration exhibits more uniformly distributed sensitivity and better small-rod contrast in the Derenzo reconstructions than H2RSPET-1CL, which are consistent with the reports in these experimental studies. It is worth mentioning that although positron range was not simulated in this work, it does not affect the spatial resolution comparison, as the positron range of \textsuperscript{18}F in a NEMA-based setup has an FWHM of approximately 0.102 mm, which is negligible relative to the system resolution.

The multi-layer extension of gPET in this work reflects deliberate design choices that prioritize computational efficiency for large-scale geometry exploration. For example, analytical blurring models for light sharing and DOI uncertainty can be adopted \citep{Geant4_doiPET_Example} rather than MC-based optical photon transport, enabling rapid and scalable evaluation of detector architectures. Recent advances in AI-based surrogate modeling, such as OptiGAN, and GPU ray-tracing engines, such as Opticks, offer promising avenues for incorporating more detailed optical modeling without incurring prohibitive computational overhead \citep{Mummaneni2025_OptiGAN, Blyth2024_Opticks}. In future, integrating gPET with next-generation optical and electronic simulation frameworks such as GATE~10 or Opticks has the potential to create an high-fidelity end-to-end PET detector design pipeline that retains the computational speed of gPET while adding extra layers of physical realism. It will be our next step to pursue these integrations and advance gPET toward a unified, high-fidelity simulation framework.

%% file: subsections/conclusion.tex
\section{Conclusion}

This work successfully extends the GPU-accelerated Monte Carlo toolkit gPET to support multi-layer PET detector geometries while maintaining its original computational efficiency and memory-optimized hierarchical description. Using this gPET extension, we evaluated a new two-layer small-animal PET design (H2RSPET-2CL) and showed that it achieved significant improvements in spatial resolution, especially at the edge of the FOV, relative to a traditional single-layer design.

%% file: subsections/acknowledgment.tex
\section*{Acknowledgment}
This work is supported in part by the National Institutes of Health (NIH) Grant 5R01EB031961.

%% file: references.bib
@article{Blyth2024_Opticks,
  author = {Blyth, Simon C.},
  title = {{Opticks}: {GPU} ray traced optical photon simulation},
  journal = {EPJ Web of Conferences},
  year = {2024},
  volume = {295},
  pages = {11014},
  doi = {10.1051/epjconf/202429511014}
}

@article{Chi2025_Review,
  author = {Chi, Yujie and Schubert, Keith and Badal, Andreu and Roncali, Emilie},
  title = {Review of {GPU}-based {Monte Carlo} simulation platforms for transmission and emission tomography in medicine},
  journal = {Physics in Medicine \& Biology},
  year = {2025},
  volume = {70},
  number = {17},
  doi = {10.1088/1361-6560/adfda7}
}

@article{Mummaneni2025_OptiGAN,
  author = {Mummaneni, Guneet and Trigila, Carlotta and Krah, Nils and Roncali, Emilie},
  title = {{optiGAN}: a deep learning-based alternative to optical photon tracking in {Python}-based {GATE (10+)}},
  journal = {Physics in Medicine \& Biology},
  year = {2025},
  note = {Accepted Manuscript},
  doi = {10.1088/1361-6560/ade2b5}
}

@article{Herraiz2024MCGPU,
  author  = {Herraiz, J. L. and Lopez-Montes, A. and Badal, A.},
  title   = {MCGPU-PET: An Open-Source Real-Time Monte Carlo PET Simulator},
  journal = {Computer Physics Communications},
  volume  = {296},
  pages   = {109008},
  year    = {2024},
  doi     = {10.1016/j.cpc.2023.109008}
}

@article{Galve2024UMC,
  author  = {Galve, Pablo and Arias-Valcayo, Fernando and Villa-Abaunza, Amaia and Ibáñez, Paula and Udías, José Manuel},
  title   = {UMC-PET: a fast and flexible Monte Carlo PET simulator},
  journal = {Physics in Medicine \& Biology},
  volume  = {69},
  number  = {3},
  pages   = {035018},
  year    = {2024},
  doi     = {10.1088/1361-6560/ad1cf9}
}

@article{Thompson2013DualLayer,
  author  = {Thompson, Christopher J. and Stortz, Greg and Goertzen, Andrew L. and Berg, Eric and Kozlowski, Piotr and Retière, Fabrice and Thiessen, Jonathan D. and Bishop, Daryl and Sossi, Vesna},
  title   = {Comparison of single and dual layer detector blocks for pre-clinical MRI-PET},
  journal = {Nuclear Instruments and Methods in Physics Research Section A: Accelerators, Spectrometers, Detectors and Associated Equipment},
  volume  = {702},
  pages   = {56-58},
  year    = {2013},
  doi     = {10.1016/j.nima.2012.07.062}
}

@article{Zhang2022DualLayer,
  author  = {Zhang, Xi and Yu, Xin and Zhu, Zhiliang and Yu, Hongsen and Zhang, Heng and Zhang, Yibin and Gu, Zheng and Xu, Jianfeng and Peng, Qiyu and Xie, Siwei},
  title   = {Development and Evaluation of a Dual-Layer-Offset PET Detector Constructed with Different Reflectors},
  journal = {Crystals},
  volume  = {12},
  number  = {1},
  pages   = {93},
  year    = {2022},
  doi     = {10.3390/cryst12010093}
}

@article{Ma2019GGEMSPET,
  author = {Ma, Bo and Gaens, M. and Caldeira, L. and Bert, J. and Lohmann, P. and Tellmann, L. and Lerche, C. and Scheins, J. and Rota Kops, E. and Xu, H. and Lenz, M. and Pietrzyk, U. and Shah, N. J.},
  title = {Scatter Correction Based on GPU-Accelerated Full Monte Carlo Simulation for Brain PET/MRI},
  journal = {IEEE Transactions on Medical Imaging},
  volume = {39},
  number = {1},
  pages = {140--151},
  year = {2020},
  note = {Epub 2019},
  doi = {10.1109/TMI.2019.2921872}
}

@article{Du2021DualEnded,
  author  = {Du, Junwei},
  title   = {Performance of Dual-Ended Readout PET Detectors Based on BGO Arrays and BaSO4 Reflector},
  journal = {IEEE Transactions on Radiation and Plasma Medical Sciences},
  volume  = {6},
  number  = {5},
  pages   = {522-528},
  year    = {2021},
  doi     = {10.1109/TRPMS.2021.3096534}
}

@article{Bouziri2021Phoswich,
  author  = {Bouziri, Haithem and Pepin, Catherine M. and Koua, Konin and Benhouria, Maher and Paulin, Caroline and Ouyang, Jinsong and Normandin, Marc and Pratte, Jean-François and El Fakhri, Georges and Lecomte, Roger and Fontaine, Réjean},
  title   = {Investigation of a Model-Based Time-over-Threshold Technique for Phoswich Crystal Discrimination},
  journal = {IEEE Transactions on Radiation and Plasma Medical Sciences},
  volume  = {6},
  number  = {4},
  pages   = {393-403},
  year    = {2021},
  doi     = {10.1109/TRPMS.2021.3077412}
}

@article{Lai2019gPET,
  author    = {Lai, Youfang and Zhong, Yiran and Chalise, A. and Shao, Y. and Jin, Mingwu and Jia, X. and Chi, Yujie},
  title     = {{gPET: a GPU-based, accurate and efficient Monte Carlo simulation tool for PET}},
  journal   = {Physics in Medicine \& Biology},
  volume    = {64},
  number    = {24},
  pages     = {245002},
  year      = {2019},
  doi       = {10.1088/1361-6560/ab5610},
  pmid      = {31711051}
}

@article{Tsuda2004_FourLayerDOI,
  author    = {Tsuda, T. and Murayama, H. and Kitamura, K. and Yamaya, T. and Yoshida, E. and Inadama, N. and Orita, N. and Haneishi, H.},
  title     = {{A four-layer depth of interaction detector block for small animal PET}},
  journal   = {IEEE Transactions on Nuclear Science},
  volume    = {51},
  number    = {5},
  pages     = {2537--2542},
  year      = {2004},
  doi       = {10.1109/TNS.2004.835777}
}

@article{Zhang2013_LORRecon,
  author    = {Zhang, Xuezhu and Stortz, Greg and Sossi, Vesna and Thompson, Christopher J. and Goertzen, Andrew L. and Qi, Jinyi},
  title     = {{Development and evaluation of a LOR-based image reconstruction with 3D system response modeling for a PET insert with dual-layer offset crystal design}},
  journal   = {Physics in Medicine \& Biology},
  volume    = {58},
  number    = {23},
  pages     = {8379--8399},
  year      = {2013},
  doi       = {10.1088/0031-9155/58/23/8379},
  pmid      = {24225131}
}

@article{Yang2008_PrototypeDOI,
  author    = {Yang, Yongfeng and Wu, Yibao and Qi, Jinyi and St. James, Sara and Du, Huini and Dokhale, Purushottam A. and Shah, Kanai S. and Farrell, Richard and Cherry, Simon R.},
  title     = {{A Prototype PET Scanner with DOI-Encoding Detectors}},
  journal   = {Journal of Nuclear Medicine},
  volume    = {49},
  number    = {7},
  pages     = {1132--1140},
  year      = {2008},
  doi       = {10.2067/jnumed.107.049791},
  pmid      = {18552140}
}

@article{Du2021_DualEnded,
  author    = {Du, Junwei},
  title     = {{Performance of Dual-Ended Readout PET Detectors Based on BGO Arrays and BaSO4 Reflector}},
  journal   = {IEEE Transactions on Radiation and Plasma Medical Sciences},
  volume    = {6},
  number    = {5},
  pages     = {522--528},
  year      = {2021},
  doi       = {10.1109/TRPMS.2021.3096534},
  pmid      = {36212107}
}

@article{Sanaat2020_MonolithicML,
  author    = {Sanaat, Amirhossein and Zaidi, Habib},
  title     = {{Depth of Interaction Estimation in a Preclinical PET Scanner Equipped with Monolithic Crystals Coupled to SiPMs Using a Deep Neural Network}},
  journal   = {Applied Sciences},
  volume    = {10},
  number    = {14},
  pages     = {4753},
  year      = {2020},
  doi       = {10.3390/app10144753}
}

@misc{Geant4_doiPET_Example,
  author    = {{The Geant4 Collaboration}},
  title     = {{Geant4 Advanced Example doiPET}},
  howpublished = {Geant4 Documentation},
  year      = {2022},
  note      = {Accessed November 9, 2025. See documentation at \url{https://geant4.web.cern.ch/docs/advanced_examples_doc/example_doiPET}}
}

@article{Jan2004GATE,
  title={{GATE: a simulation toolkit for PET and SPECT}},
  author={Jan, S. and Santin, G. and Strul, D. and Staelens, S. and Assi{\'e}, K. and Autret, D. and Avner, S. and Barbier, R. and Bardi{\`e}s, M. and Bloomfield, P.M. and others},
  journal={Physics in Medicine \& Biology},
  volume={49},
  number={19},
  pages={4543--4561},
  year={2004},
  doi={10.1088/0031-9155/49/19/007}
}

@article{zeng2024depth,
  title={Depth-encoding using optical photon TOF in a prism-PET detector with tapered crystals},
  author={Zeng, Xinjie and LaBella, Andy and Wang, Zipai and Li, Yixin and Tan, Wanbin and Goldan, Amir H},
  journal={Medical physics},
  volume={51},
  number={6},
  pages={4044--4055},
  year={2024},
  publisher={Wiley Online Library}
}

@article{shukla2006positron,
  title={Positron emission tomography: An overview},
  author={Shukla, Arvind K and Kumar, Utham},
  journal={Journal of medical physics},
  volume={31},
  number={1},
  pages={13--21},
  year={2006},
  publisher={Medknow}
}

@article{soret2007partial,
  title={Partial-volume effect in PET tumor imaging},
  author={Soret, Marine and Bacharach, Stephen L and Buvat, Irene},
  journal={Journal of nuclear medicine},
  volume={48},
  number={6},
  pages={932--945},
  year={2007},
  publisher={Society of Nuclear Medicine}
}

@article{merlin2018castor,
  title={CASToR: a generic data organization and processing code framework for multi-modal and multi-dimensional tomographic reconstruction},
  author={Merlin, Thibaut and Stute, Simon and Benoit, Didier and Bert, Julien and Carlier, Thomas and Comtat, Claude and Filipovic, Marina and Lamare, Fr{\'e}d{\'e}ric and Visvikis, Dimitris},
  journal={Physics in Medicine \& Biology},
  volume={63},
  number={18},
  pages={185005},
  year={2018},
  publisher={IOP Publishing}
}

@article{Liu2022,
  author    = {Zheng Liu and Ming Niu and Zhonghua Kuang and Ning Ren and San Wu and Longhan Cong and Xiaohui Wang and Ziru Sang and Crispin Williams and Yongfeng Yang},
  title     = {High resolution detectors for whole-body PET scanners by using dual-ended readout},
  journal   = {EJNMMI Physics},
  year      = {2022},
  volume    = {9},
  number    = {1},
  pages     = {29},
  doi       = {10.1186/s40658-022-00460-4}
}
